\documentclass[seceq,supplement]{ptptex}
\usepackage{epsfig}
\usepackage{graphicx}
\usepackage{latexsym}

\title{Finite Density Lattice Gauge Theories with Positive Fermion Determinants
\footnote{Talk presented by D.~K.~Sinclair at {\it Finite Density QCD at Nara},
Nara, Japan, 10th-12th July, 2003.}}
\author{D.~K.~Sinclair$^1$ \\ J.~B.~Kogut$^2$ and D.~Toublan$^2$}
\inst{$^1$HEP Division, Argonne National Laboratory, 9700 South Cass Avenue,
      Argonne, IL 60439, USA \\
      $^2$Department of Physics, University of Illinois, 1110 West Green Street,
      Urbana, IL 61801, USA}
\abst{We perform simulations of (3-colour) QCD with 2 quark flavours at a
finite chemical potential $\mu_I$ for isospin($I_3$), and of 2-colour QCD at a
finite chemical potential $\mu$ for quark number. At zero temperature, QCD at
finite $\mu_I$ has a mean-field phase transition at $\mu_I=m_\pi$ to a
superfluid state with a charged pion condensate which spontaneously breaks
$I_3$. We study the finite temperature transition as a function of $\mu_I$.
For $\mu_I < m_\pi$, where this is closely related to the transition at finite
$\mu$, this appears to be a crossover independent of quark mass, with no sign
of the proposed critical endpoint. For $\mu_I > m_\pi$ this becomes a true
phase transition where the pion condensate evaporates. For $\mu_I$ just above
$m_\pi$ the transition seems to be second order, while for larger $\mu_I$ it
appears to become first order. At zero temperature, 2-colour QCD also possesses
a superfluid state with a diquark condensate. We study its spectrum of
Goldstone and pseudo-Goldstone bosons associated with chiral and quark-number
symmetry breaking.}

\begin{document}

\maketitle

\section{Introduction}

We are interested in QCD at finite baryon-number, isospin and strangeness
densities at zero and at finite temperature. Finite baryon-number is of most
interest, but most difficult to simulate. QCD at finite chemical potential 
$\mu$ for quark number has a complex fermion determinant which prevents the
use of standard simulation methods.

We are studying theories which possess {\it some} of the properties of QCD at 
finite $\mu$, but have positive fermion determinants. This allows us to use
standard (hybrid molecular-dynamics) simulations.

2-colour QCD at finite $\mu$ has a positive fermion pfaffian and is amenable
to simulations \cite{Hands:1999md,Kogut:2001na}. 
It has a sensible meson spectrum, but unphysical `baryon'
spectrum. At $\mu=m_\pi/2$, it undergoes a phase transition  to a superfluid 
state with a diquark condensate and Goldstone bosons 
\cite{Kogut:1999iv,Kogut:2000ek,Splittorff:2001fy}. 
This is the analogue of the
colour superconducting state suggested for QCD at finite $\mu$. We have also
studied this theory at non-zero temperature \cite{Kogut:2001if}. Here we
summarize our calculation of the spectrum of (pseudo)-Goldstone bosons which
characterize the pattern of symmetry breaking for this theory 
\cite{Kogut:2003ju}.

QCD at a finite chemical potential $\mu_I$ for isospin ($I_3$), has a positive
fermion determinant which allows simulations \cite{Kogut:2002zg}. 
It undergoes a phase transition
at $\mu_I=m_\pi$ to a superfluid state with a charged pion condensate which
breaks $I_3$ and parity spontaneously, with associated charged Goldstone pions
\cite{Son:2000xc,Son:2000by}. 

\begin{figure}[htb]
\parbox{\halftext}{\includegraphics[width=6.6cm]{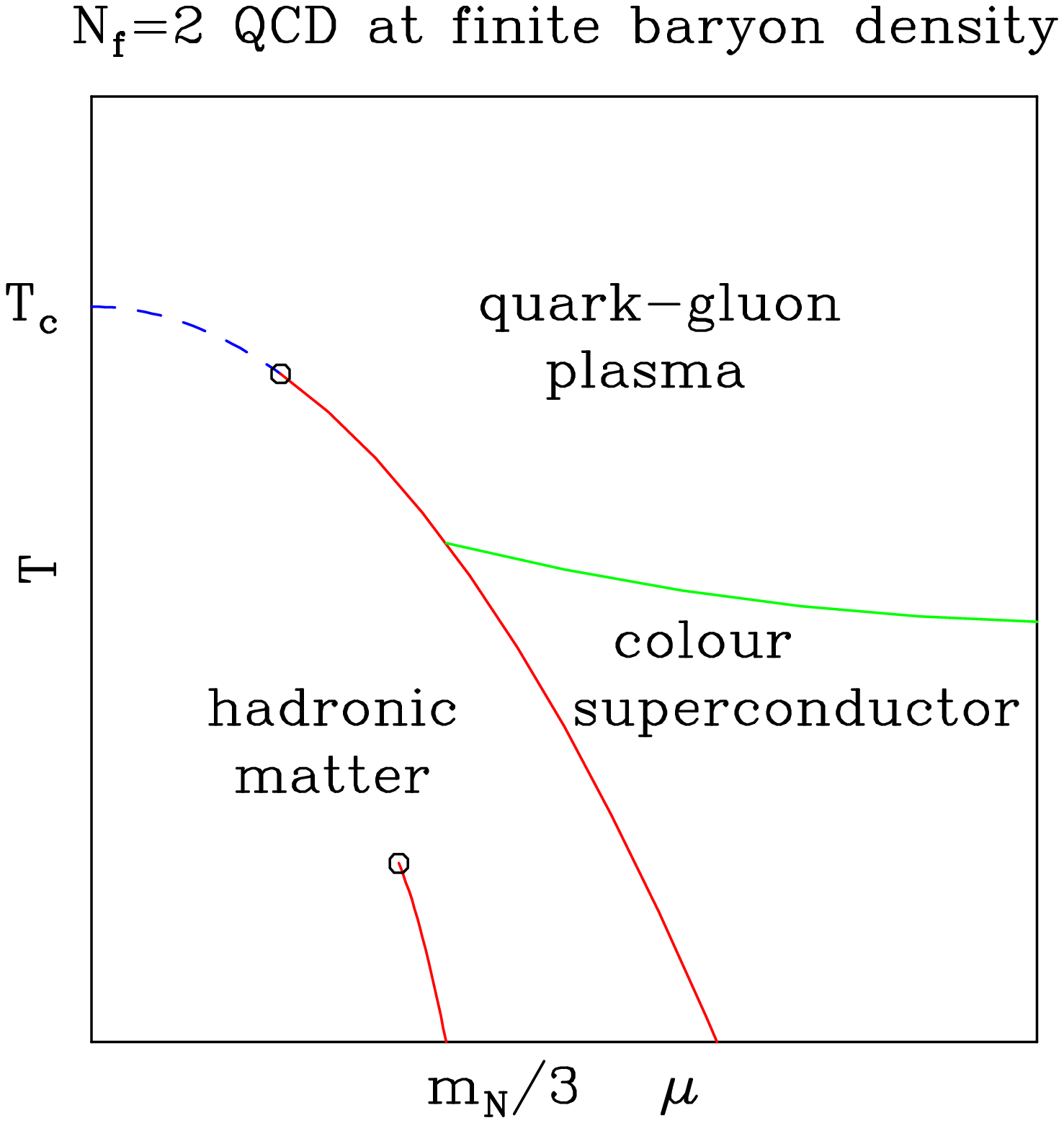}}
\hfill
\parbox{\halftext}{\includegraphics[width=6.6cm]{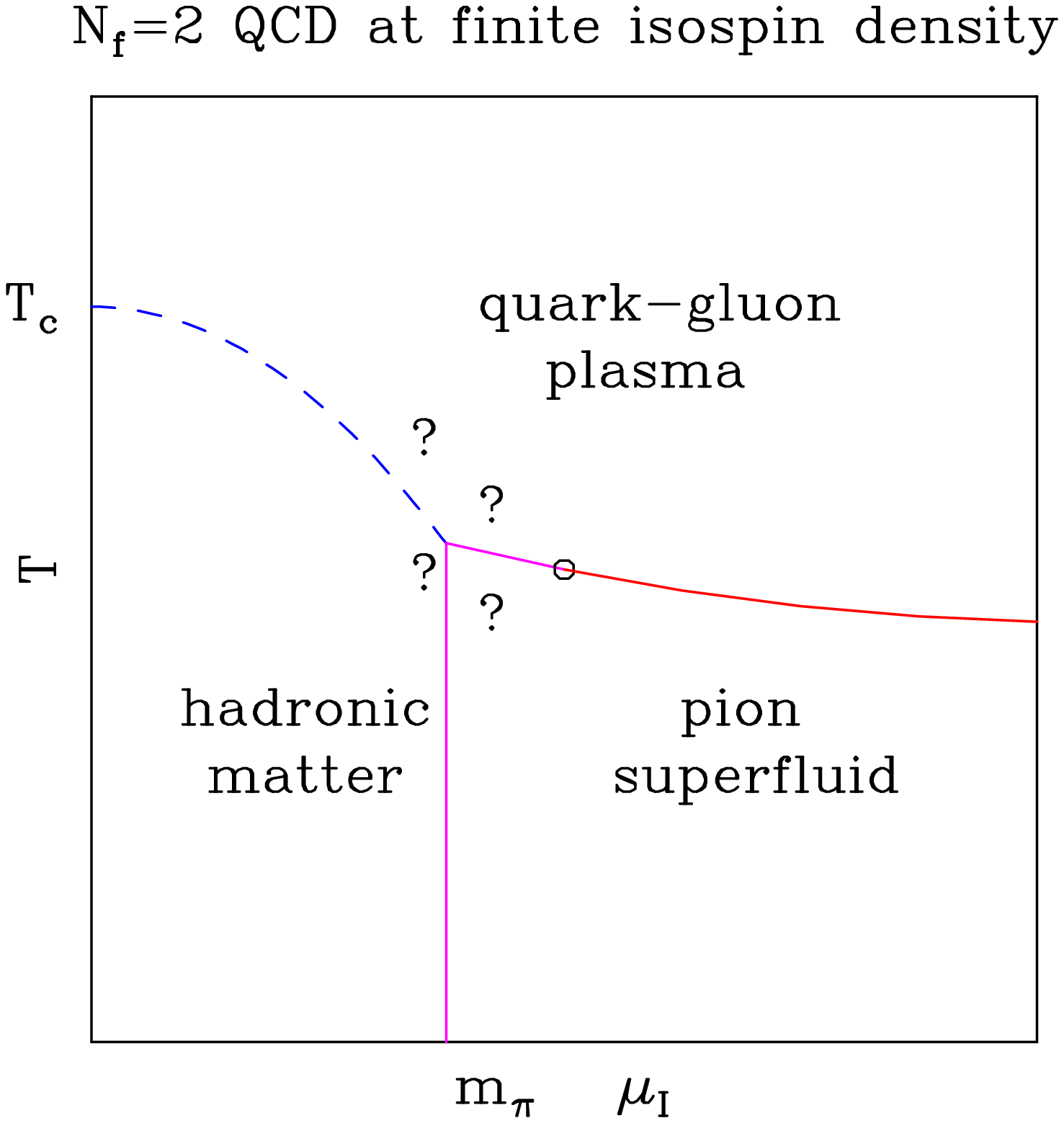}}
\caption{a) Proposed phase diagram for QCD at finite quark-number chemical
potential $\mu$ and temperature $T$. b)Proposed phase diagram for QCD at 
finite isospin chemical potential $\mu_I$ and temperature $T$.}
\label{fig:phase}
\end{figure}

Let us now compare the proposed phase diagrams for QCD at finite $\mu$ 
\cite{Shuryak:1999gi,Alford:2001dt} and
QCD at finite $\mu_I$ (figure \ref{fig:phase}). Both show crossovers (dashed
lines) at small chemical potential. For QCD at finite $\mu$ \ref{fig:phase}a
the 2 solid lines which intersect the $T=0$ axis and end in open circles
(critical points) are expected to be first order. The order of the transition
between the colour-superconducting and quark-gluon plasma phases is unknown.
For QCD at finite $\mu_I$, the solid lines to the left of the open circle
(tricritical point) are second order, while that to the right is first order.

We are studying 2-flavour QCD at finite $\mu_I$ and temperature in the
neighbourhood of the finite temperature transition 
\cite{Kogut:2002se,Kogut:2003cd}. 
At low $\mu_I$ this
transition from hadronic matter to a quark-gluon plasma is related to the
corresponding transition at finite $\mu$. Our predictions for the small $\mu$
behaviour of the finite temperature transition can be compared directly with
those obtained by de Forcrand and Philipsen by continuing from imaginary $\mu$
\cite{deForcrand:2002ci}.
These results  will also be compared with results obtained using the series 
methods of the Bielefeld-Swansea collaboration 
\cite{Allton:2002zi,Allton:2003vx}. 
They could also be checked using the reweighting methods of Fodor and Katz
\cite{Fodor:2001au,Fodor:2001pe,Fodor:2002km}.
We are also simulating this theory at large $\mu_I$ where the finite
temperature transition occurs at the point where the pion condensate
evaporates. It is therefore a true phase transition. In particular we wish to
determine where this transition changes from second order to first
\cite{Kogut:2002zg}.

Section~2 gives our results for lattice QCD at finite $\mu_I$ and zero 
temperature. In Section~3 we present preliminary results for QCD at finite
$\mu_I$ and temperature. The (pseudo)-Goldstone spectrum for 2-colour lattice
QCD at finite $\mu$ is given in section~4. In section~5 we indicate how our
new $\chi$QCD action might help overcome some of the problems at finite 
$\mu/\mu_I$. Section~6 gives our conclusions and suggests future directions for 
investigation.

\section{Lattice QCD at finite $\mu_I$}

The staggered quark action for lattice QCD at finite $\mu_I$ is
\begin{equation}
S_f=\sum_{sites} \left[\bar{\chi}[D\!\!\!\!/(\frac{1}{2}\tau_3\mu_I)+m]\chi
                   + i\lambda\epsilon\bar{\chi}\tau_2\chi\right]
\end{equation}
where $D\!\!\!\!/(\frac{1}{2}\tau_3\mu_I)$ is the standard staggered quark
$D\!\!\!\!/$ with the links in the $+t$ direction multiplied by 
$\exp(\frac{1}{2}\tau_3\mu_I)$ and those in the $-t$ direction multiplied by
$\exp(-\frac{1}{2}\tau_3\mu_I)$. The $\lambda$ term is an explicit $I_3$ 
symmetry breaking term required to see spontaneous symmetry breaking on a 
finite lattice. The determinant
\begin{equation}
\det[D\!\!\!\!/(\frac{1}{2}\tau_3\mu_I) + m + i\lambda\epsilon\tau_2]
               =\det[{\cal A}^\dagger{\cal A}+\lambda^2],
\end{equation}
where
\begin{equation}
{\cal A} \equiv D\!\!\!\!/(\frac{1}{2}\mu_I)+m ,
\end{equation}
is positive allowing us to use standard hybrid molecular dynamics simulations.

This lattice action has a global $U(2) \times U(2)$ flavour symmetry when
$m=\mu_I=\lambda=0$. This breaks spontaneously to $U(2)$, with 4 Goldstone 
pions. When $\mu_I=0$ but $m \ne 0$, this is reduced to $U(2)_V$, and the 
would-be-Goldstone pions gain a mass by PCAC. If $m=0$, but $\mu_I \ne 0$, the 
symmetry is reduced to $U(1) \times U(1) \times U(1) \times U(1)$. Finally, if 
$m \ne 0$ and $\mu_I \ne 0$, the symmetry is further reduced to 
$U(1)_V \times U(1)_V$.

As $\mu_I$ is increased from zero, the effective mass of the $\pi^{+}$ is
reduced to
\begin{equation}
m_\pi(\mu_I) = m_\pi-\mu_I,
\end{equation}
which vanishes at $\mu_I=m_\pi$. For $\mu_I \ge m_\pi$, the $U(1)_V$ symmetry
associated with $\tau_3$ is broken spontaneously by a charged pion condensate
\begin{equation}                                                              
i\langle\bar{\chi}\epsilon\tau_2\chi\rangle \Leftrightarrow
i\langle\bar{\psi}\gamma_5\tau_2\psi\rangle.
\end{equation}                                                                
The charged pion excitation created by%
\begin{equation}
i\bar{\chi}\epsilon\tau_1\chi
\end{equation}                                             
is the associated Goldstone boson. When $\lambda$ is non-zero this gains a mass.
Note that if $m=0$, there will actually be 2 Goldstone bosons.

\begin{figure}[htb]
\parbox{\halftext}{\includegraphics[width=6.6cm]{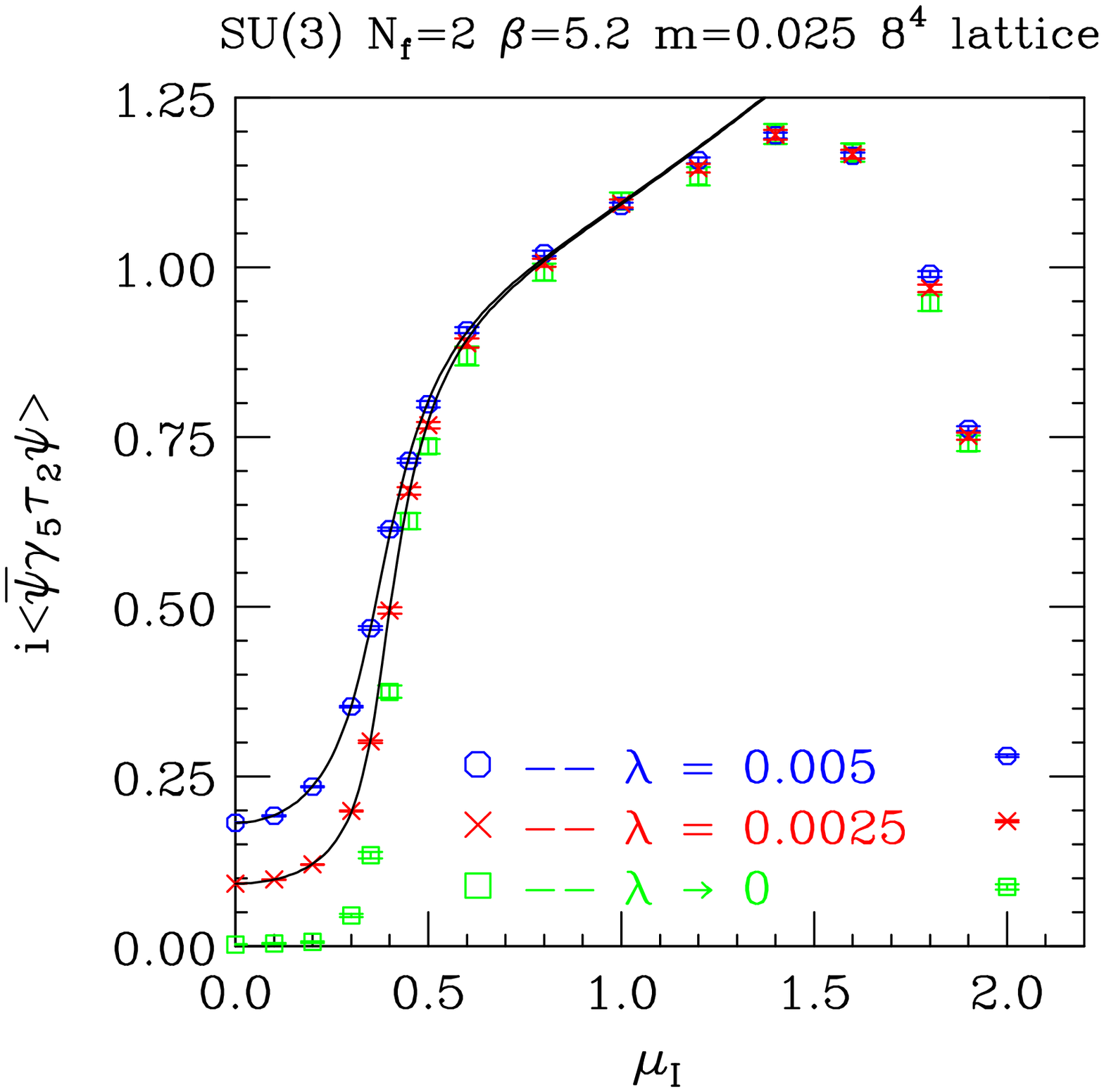}}
\hfill
\parbox{\halftext}{\includegraphics[width=6.6cm]{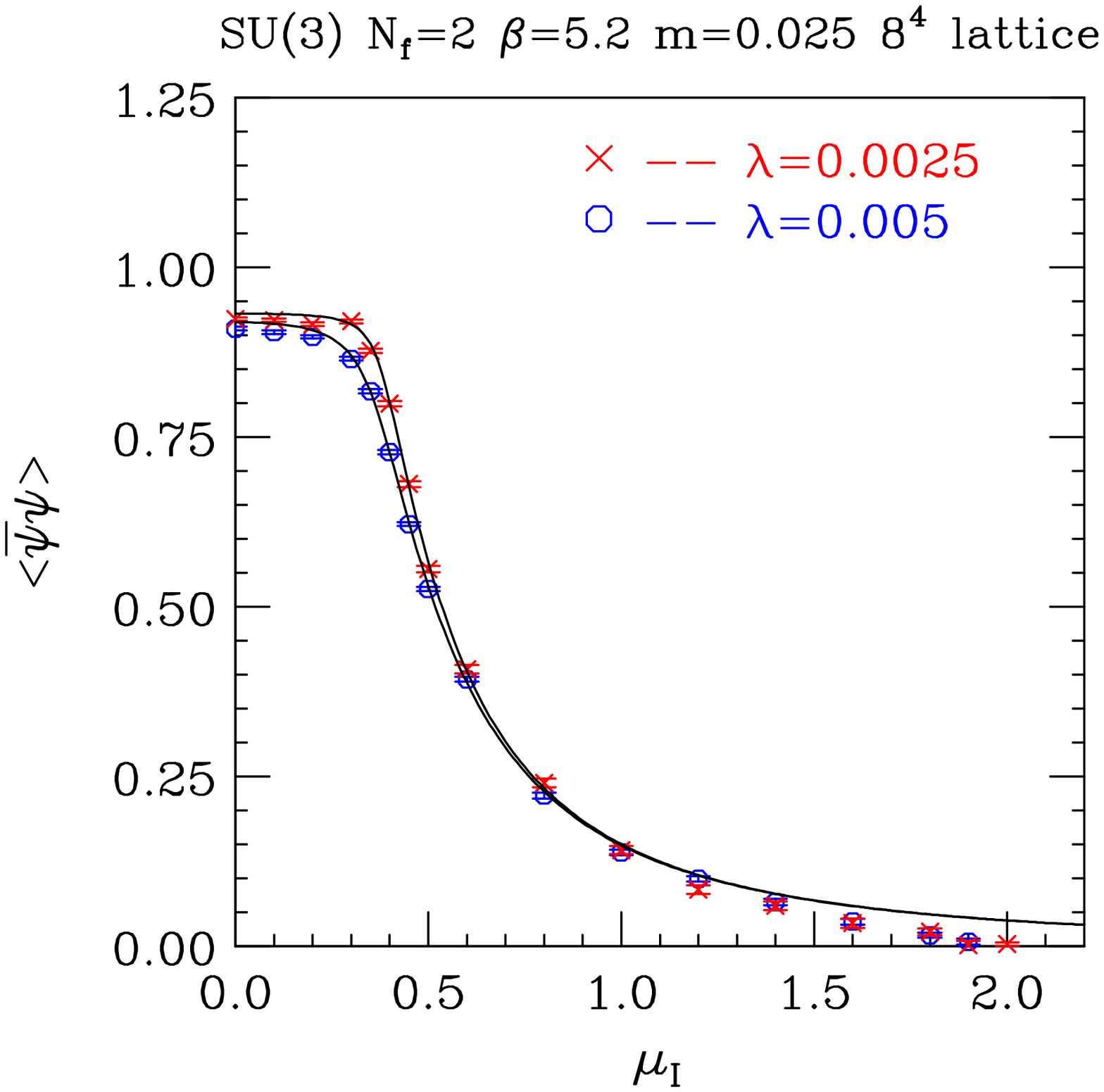}}
\caption{a) Charged pion condensate as a function of $I_3$ chemical
         potential $\mu_I$. The curves are a fit to mean-field scaling.
         b) Chiral condensate as a function of chemical potential$\mu_I$. 
         The curves are predictions from the fits to the pion condensate.}
\label{fig:condensates}
\end{figure}

Simulations of lattice QCD at finite $\mu_I$ were performed with $m=0.025$ and 
$\lambda=0.0025, 0.005$ on an $8^4$ lattice \cite{Kogut:2002zg}. 
Figure~\ref{fig:condensates} shows the diquark and
chiral condensates as functions of $\mu_I$. 
For $m_\pi$ and $\mu_I$ small one can use chiral perturbation theory to 
predict the behaviour of this theory \cite{Son:2000xc,Son:2000by}. 
At tree-level this shows the phase
transition at $\mu_I=m_\pi$. For $\mu_I > m_\pi$, the condensate rotates from
the chiral direction towards the pion condensate direction so that $I_3$
breaks spontaneously. Indications are that at higher orders, the transition
remains at $\mu_I=m_\pi$, but the rotation of the condensate is accompanied
by an increase of its magnitude. However, the transition remains mean-field.
We model this behaviour by fitting to a scaling form based on the tree-level
predictions of a chiral Lagrangian of the linear sigma model form which allows
the magnitude of the condensate to vary, while retaining the other predictions
of chiral perturbation theory, with only one additional fitting parameter.
The fits are shown on the figures. These indicate that there is a phase 
transition with mean-field exponents to a state where $I_3$ and parity are
broken spontaneously. These fits also predict how the $I_3$ density 
increases from zero above $m_\pi$, and are in accord with the data (not shown)
provided $\mu_I$ is not too large.

\section{Lattice QCD at finite $\mu_I$ and $T$}

We first consider the finite temperature transition for $\mu_I < \mu_c$. As
$\mu_I$ is increased from zero, the $\beta_c$ of the crossover from hadronic
matter to a quark gluon plasma decreases. Note, the results in this section
should be considered preliminary. 

We have performed simulations on an $8^3 \times 4$ lattice at $m=0.05$ and
$\lambda=0$, at values of $\mu_I$ up to $0.55$, which is very close to $\mu_c$
\cite{Kogut:2002se,Kogut:2003cd}.
The position of the crossover is determined from the maxima of the 
susceptibilities for the plaquette, the chiral condensate, the Wilson line
(Polyakov Loop) and the isospin density. As we shall see, there is consistency
between these estimates. (The close proximity of the transitions in these
various observables has been noted by many authors. Recently attempts to
understand the reason for this have been made for zero $\mu/\mu_I$
\cite{Hatta:2003ga}. Combining this with earlier observations for finite $\mu$
\cite{Fukushima:2002mp}, it seems probable that this argument works at finite
$\mu$ (and hence $\mu_I$). A detailed analysis in terms of effective field
theories, which also indicates applicability to finite $\mu/\mu_I$, has been
given in \cite{Mocsy:2003tr,Mocsy:2003qw}$\,$.)
We use Ferrenberg-Swendsen reweighting  
\cite{Ferrenberg:yz} to determine
these values with some precision. For the chiral condensate and isospin
density which are determined from stochastic estimators, we use 5 noise vectors
to calculate the susceptibilities, discarding the diagonal terms to obtain an
unbiased estimate.

\begin{figure}[htb]
\parbox{\halftext}{\includegraphics[width=6.6cm]{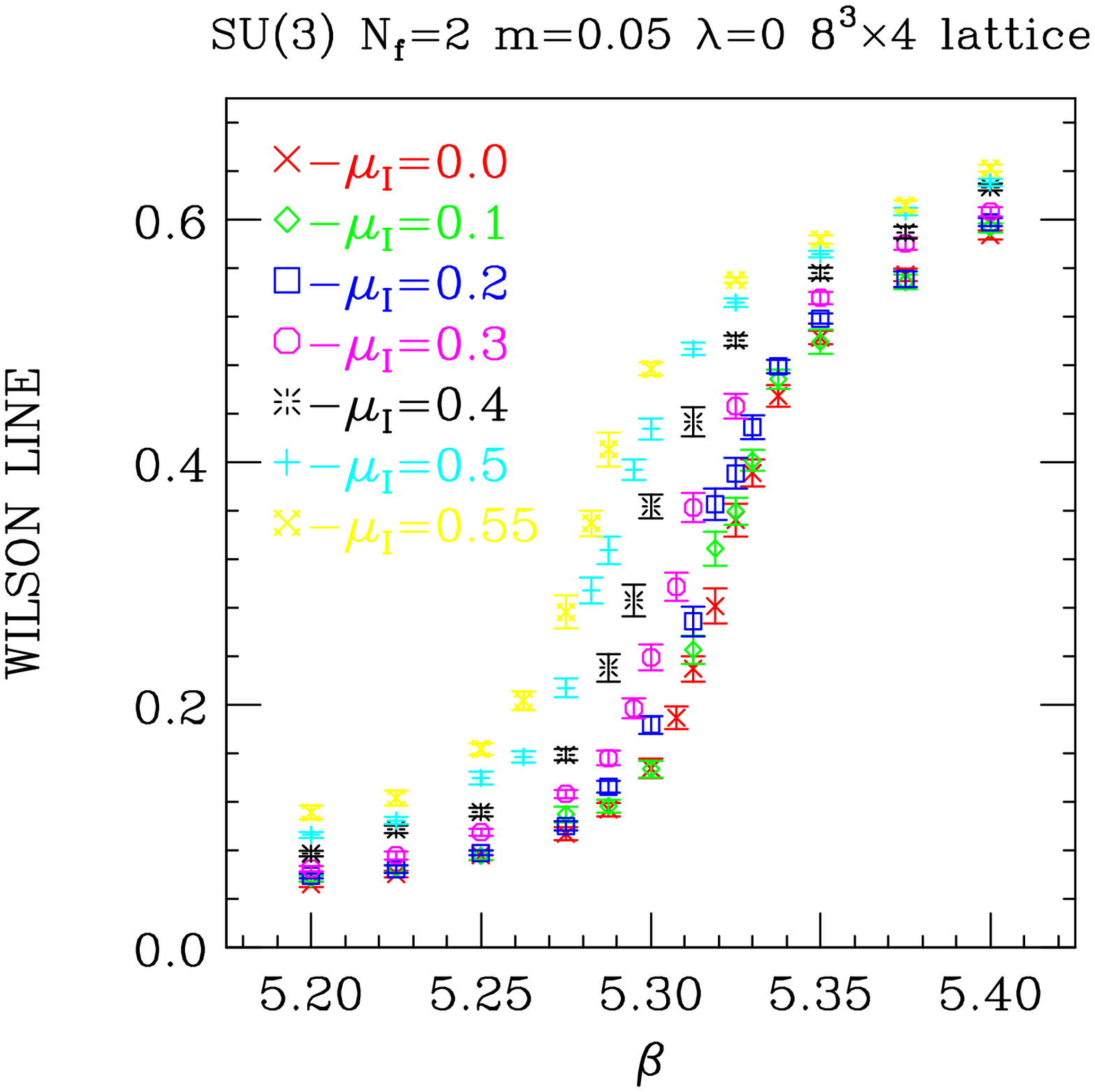}
\caption{Wilson Line as a function of $\beta=6/g^2$ for $\mu_I$ values 
$\le 0.55$.}\label{fig:wilson0.05}}   
\hfill
\parbox{\halftext}{\includegraphics[width=6.6cm]{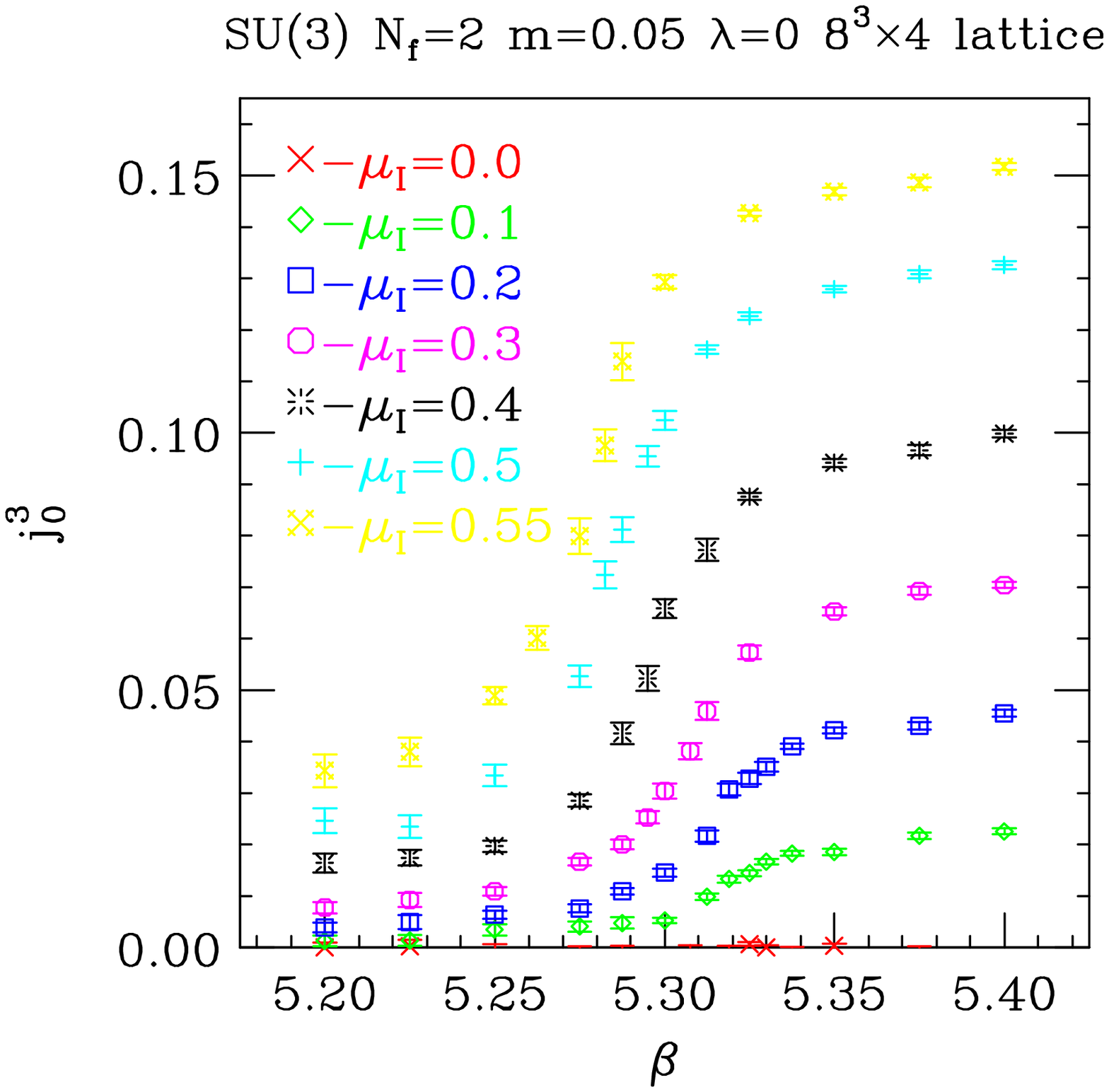}
\caption{Isospin density as a function of $\beta=6/g^2$ for $\mu_I$ values 
$\le 0.55$.}\label{fig:j0_0.05}}
\end{figure}

In figure~\ref{fig:wilson0.05} we show the Thermal Wilson Line (Polyakov Loop)
as a function of $\beta=6/g^2$ for a selection of $\mu_I$ values in this
range. First is is clear that for each $\mu_I$ there is a rapid crossover from
hadronic matter to a quark-gluon plasma over an interval 
$\Delta\beta \sim 0.05$. What we also notice is that the $\beta$ of the
crossover decreases monotonically with increasing $\mu_I$.

The corresponding graph for the $I_3$ density is given in 
figure~\ref{fig:j0_0.05}. Here, not only does the $\beta$ value for the cross
over decrease with increasing $\mu_I$, the value of this density increases
with increasing $\mu_I$. This is expected, since when $\mu_I$ is increased, it
raises the level of the fermi sea, and hence the excess of $I_3=1/2$ quarks 
over $I_3=-1/2$ quarks.

\begin{figure}[htb]
\parbox{\halftext}{\includegraphics[width=6.6cm]{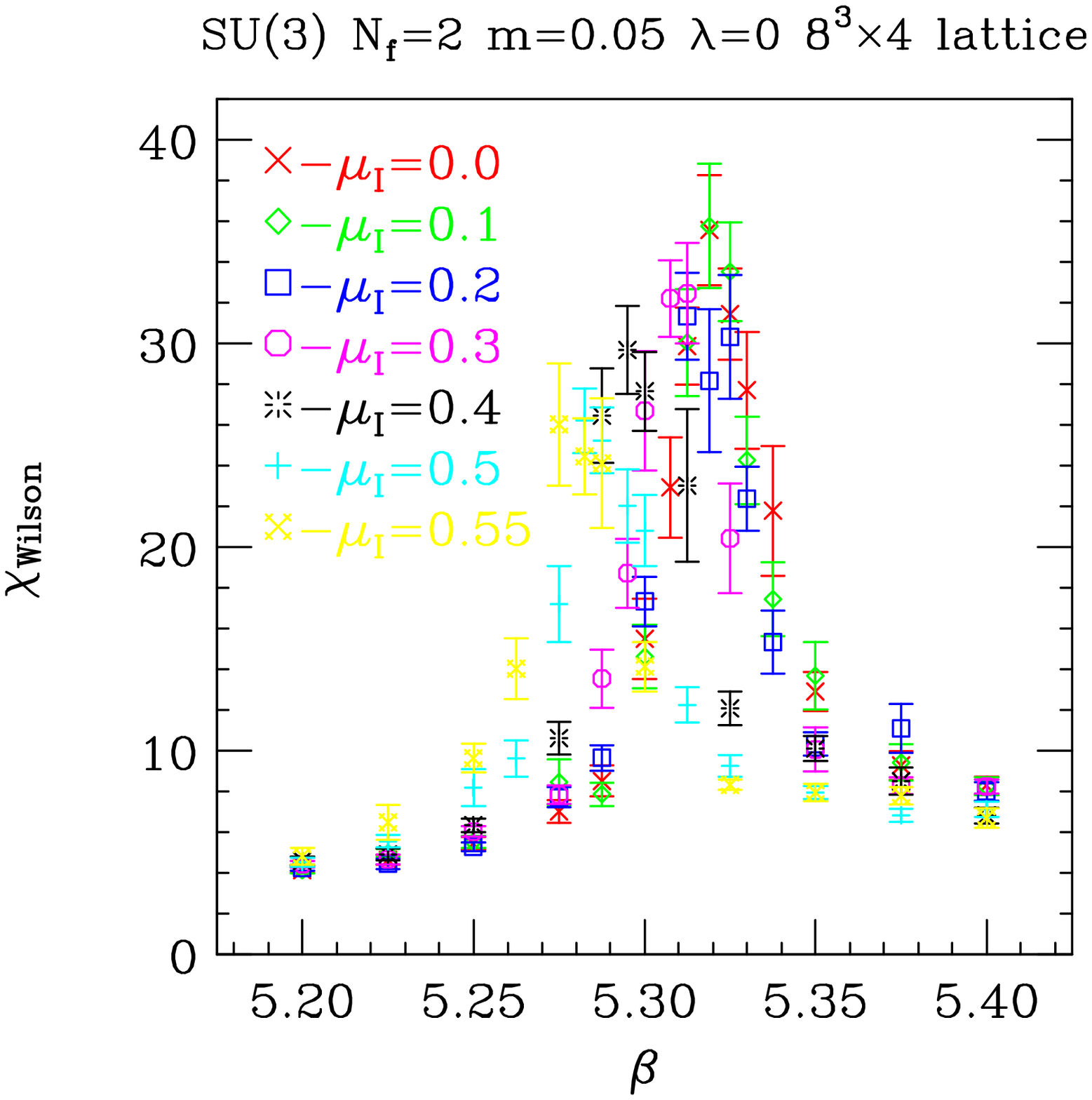}
\caption{Wilson Line susceptibility as a function of $\beta=6/g^2$ for
$\mu_I$ values $\le 0.55$.}\label{fig:suscwl}}
\hfill
\parbox{\halftext}{\includegraphics[width=6.6cm]{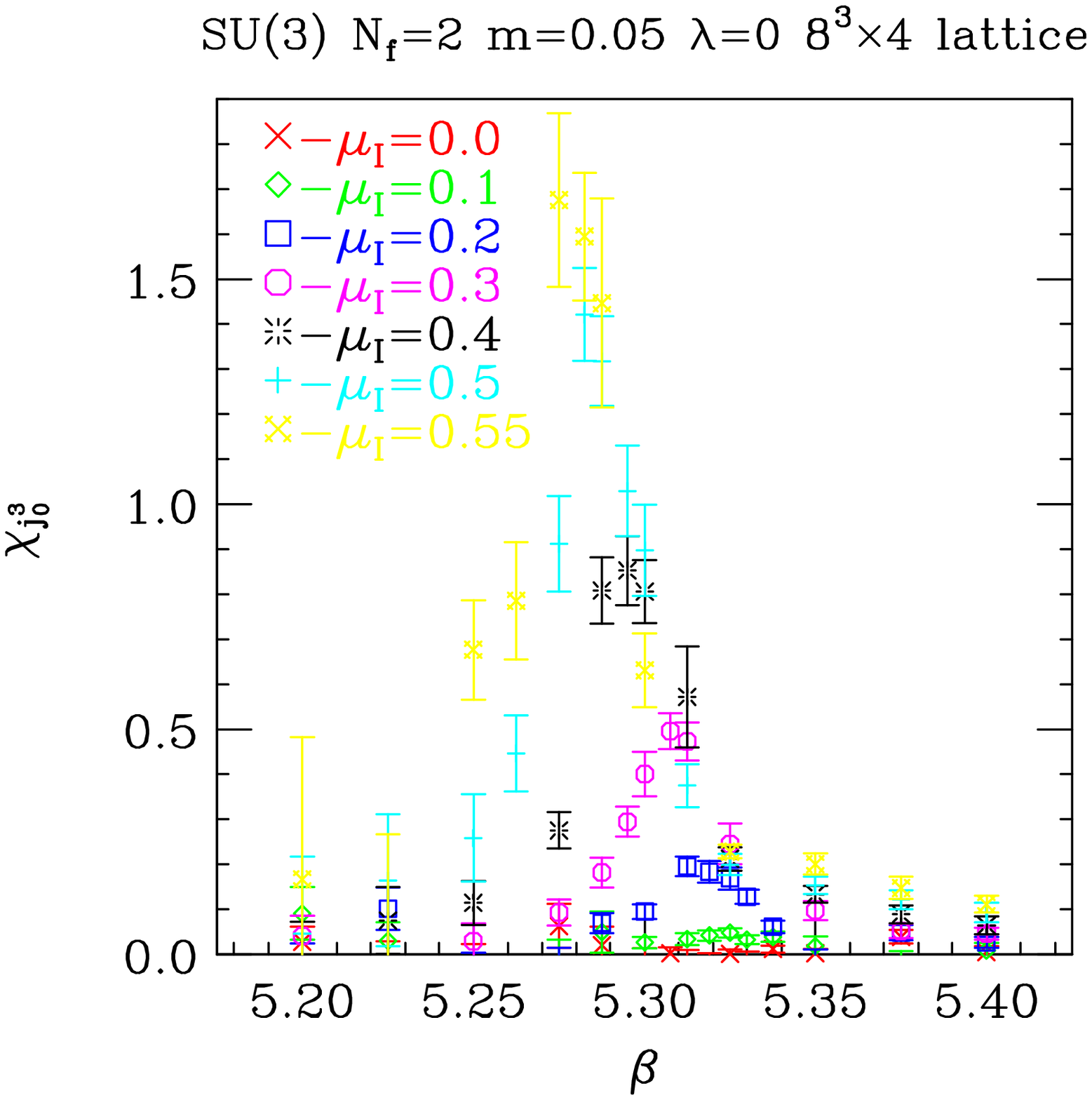}
\caption{Isospin susceptibility as a function of $\beta=6/g^2$ for
$\mu_I$ values $\le 0.55$.}\label{fig:suscj0}}
\end{figure}

In figure~\ref{fig:suscwl} we show the Wilson Line susceptibility and in
figure~\ref{fig:suscj0} we show the susceptibility for the isospin density.
The chiral and plaquette susceptibilities are similar to that for the Wilson
Line. The susceptibility for an operator ${\cal O}$ is defined as
\begin{equation}
\chi_{\cal O} = V \langle {\cal O}^2 - \langle{\cal O}\rangle^2 \rangle,
\end{equation}
where $V$ is the space-time volume of the lattice.
The peak of the susceptibility serves to define the position of the crossover.
Clearly the measurements in these plots are inadequate to give precise
positions of the peaks. However we can make a Ferrenberg-Swendsen reweighting of
our `data' at $\beta$ values close to the peaks, which yields a more precise
estimate of their positions.

\begin{figure}[htb]
\centerline{\includegraphics[width=10cm]{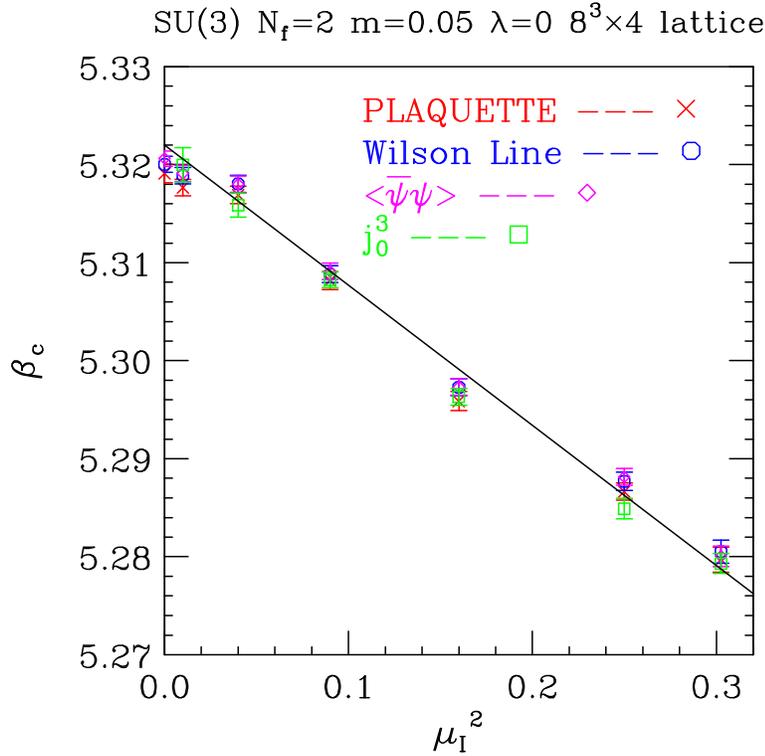}}
\caption{$\beta_c$ determined from the maxima of the susceptibilities,
as a function of $\mu_I^2$.}
\label{fig:beta_c}
\end{figure}

We plot our estimates of $\beta_c$ versus $\mu_I^2$ from these 4 
susceptibilities in figure~\ref{fig:beta_c}, since we expect the leading 
dependence to be quadratic in $\mu_I$.
The straight line in this graph 
\begin{equation}
\beta_c=5.322-0.143\mu_I^2,
\label{eqn:beta_c}
\end{equation}
is not a proper fit
and is only meant to be a rough guide. If we assume that at $\mu_I=0$, $T_c
\approx 173$~MeV, then 2-loop running of the coupling would predict that for
the highest $\mu_I$ considered, $T_c(\mu_I=0.55) \approx 164$~MeV. Hence the
relatively weak dependence of $\beta_c$ on $\mu_I$ does translate into a
relatively small change in $T_c$ over a fairly large range of $\mu_I$.

The Swansea-Bielefeld collaboration have determined that, for quark-number
chemical potential $\mu$ small enough, the phase of the fermion determinant
is well enough behaved that simulating with the magnitude of the determinant
and including the phase factor in the measurements should work. This means
that, to the extent that the position of the crossover is a well-defined 
quantity (i.e. is the same for all observables), the dependence of $T_c$ on
$\mu$ and $\mu_I$ should be identical for small $\mu$, in particular,
\begin{equation}
\beta_c(\mu)=\beta_c(\mu_I=2\mu)
\label{eqn:mu/mu_I}
\end{equation}
so we can predict the small $\mu$ behaviour of $\beta_c$ and hence $T_c$. Our
results in equation~\ref{eqn:beta_c}, using equation~\ref{eqn:mu/mu_I},
appear to be in agreement with those of
de Forcrand and Philipsen \cite{deForcrand:2002ci}

This also suggests that we should look for a critical point where the crossover
becomes a first order transition, since such a critical endpoint is predicted
for QCD at a finite chemical potential $\mu$ for quark number. For the
$m=0.05$ simulations reported here, we see no evidence for this for 
$\mu_I \le 0.55$ (i.e. for $\mu_I < \mu_c$). However, $\mu_I=0.55$ is
equivalent to $\mu=0.275 \approx 180$~MeV ($\mu_B=535$~MeV), which is less
than $\mu$ measured for the critical point by Fodor and Katz
\cite{Fodor:2001au,Fodor:2001pe,Fodor:2002km}. (Note, however,
that the  Fodor-Katz simulations are for $2+1$-flavours, not $2$.) We are
therefore performing simulations with $m=0.1$ and $m=0.2$ where $\mu_c$ is
beyond the Fodor-Katz value. However, we still see no signals (so far) for
first-order behaviour. Thus we conclude that if there is a transition to be
identified with this critical point, its $\mu_I$ increases with increasing
mass, which is not unexpected.

Now let us turn to our studies of the finite temperature transition for larger
$\mu_I$s. For $\mu_I > \mu_c$, the pion condensate evaporates at the finite
temperature transition. Thus here the finite temperature transition {\it is} a 
phase transition. At low temperatures, by approaching this phase boundary in 
the $\mu_I$ direction we have determined that the transition is second order. 
We now have preliminary evidence that for $m=0.05$, $\lambda=0.005$ and 
$N_t=4$, the $\mu_I=0.8$ transition is first order, based on simulations on 
$8^3 \times 4$ and $16^3 \times 4$ lattices, and are in the process of 
determining the position of the tricritical point. 

\begin{figure}[htb]
\parbox{\halftext}{\includegraphics[width=6.6cm]{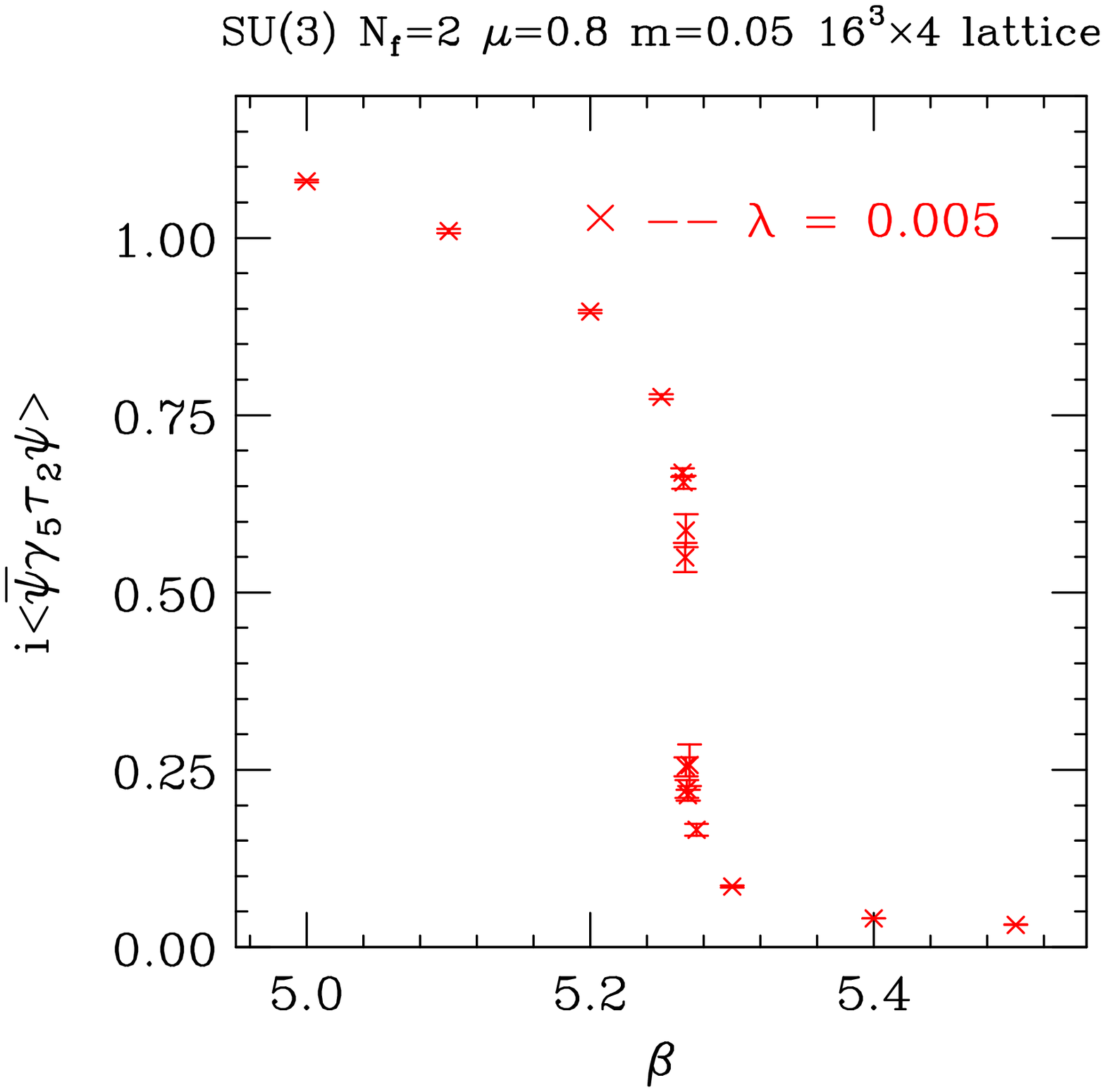}}
\hfill
\parbox{\halftext}{\includegraphics[width=6.6cm]{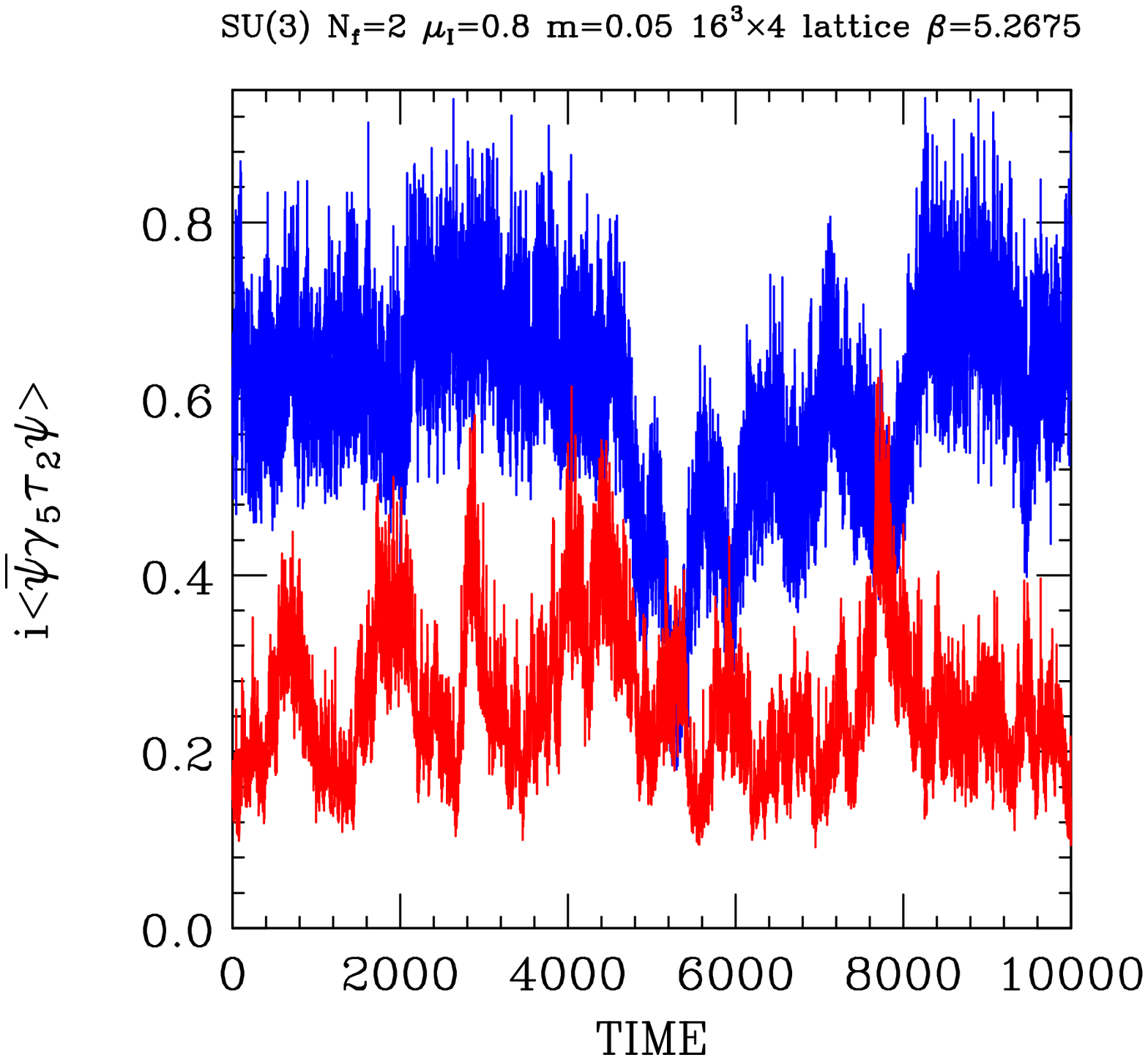}}
\caption{a) Pion condensate as a function of $\beta=6/g^2$ for $m=0.05$ and
$\mu_I=0.8$. b) Pion condensates as functions of molecular-dynamics time at
$\beta=5.2675$ from hot and cold starts.}
\label{fig:pg5t2p}
\end{figure}

We note that this transition is at $\beta_c \approx 5.2675$ i.e. at 
$T_c \approx 162$~MeV. This $\mu_I$ value is equivalent to 
$\mu=0.4 \approx 259$~MeV. Hence the tricritical point must be close to
the Fodor-Katz critical endpoint where $T_c \approx 160$~MeV, 
$\mu \approx 242$~MeV. We can therefore speculate that these 2 transitions are
somehow related. Figure~\ref{fig:pg5t2p}a shows the pion condensate as a
function of $\beta$ on a $16^3 \times 4$ lattice at $\mu=0.8$. 
Figure~\ref{fig:pg5t2p}b shows the time evolution of this condensate from
hot and cold starts at $\beta=5.2675$, showing evidence for coexisting states.
Our ``time'' increment for updating $dt=0.02$ was larger than desirable so we
are repeating our simulations with $dt=0.01$ close to the transition. 

\section{The Spectrum of 2-colour QCD}

Lattice 2-colour QCD at finite chemical potential $\mu$ for quark number has 
the staggered fermion action
\begin{equation}
S_f = \sum_{sites}\left\{\bar{\chi}[D\!\!\!\!/\,(\mu) + m]\chi
      + \frac{1}{2}\lambda[\chi^T\tau_2\chi + 
                        \bar{\chi}\tau_2\bar{\chi}^T]\right\}
\label{eqn:lagrangian}
\end{equation}
This theory has a positive pfaffian which allows us to simulate it using the
hybrid molecular-dynamics method. This lattice action has a $U(2)$ flavour
symmetry when $m=\mu=\lambda=0$. When this breaks spontaneously 
$U(2) \rightarrow U(1)$ giving rise to 3 Goldstone bosons. We have studied the
behaviour of these bosons when these parameters are no longer zero. In
particular we study the $\mu$ dependence when $m$ is fixed at a value small
enough to be within reach of chiral perturbation theory, and $\lambda << m$
\cite{Kogut:2001na,Kogut:2003ju}.

\begin{figure}[htb]
\parbox{\halftext}{\includegraphics[width=6.6cm]{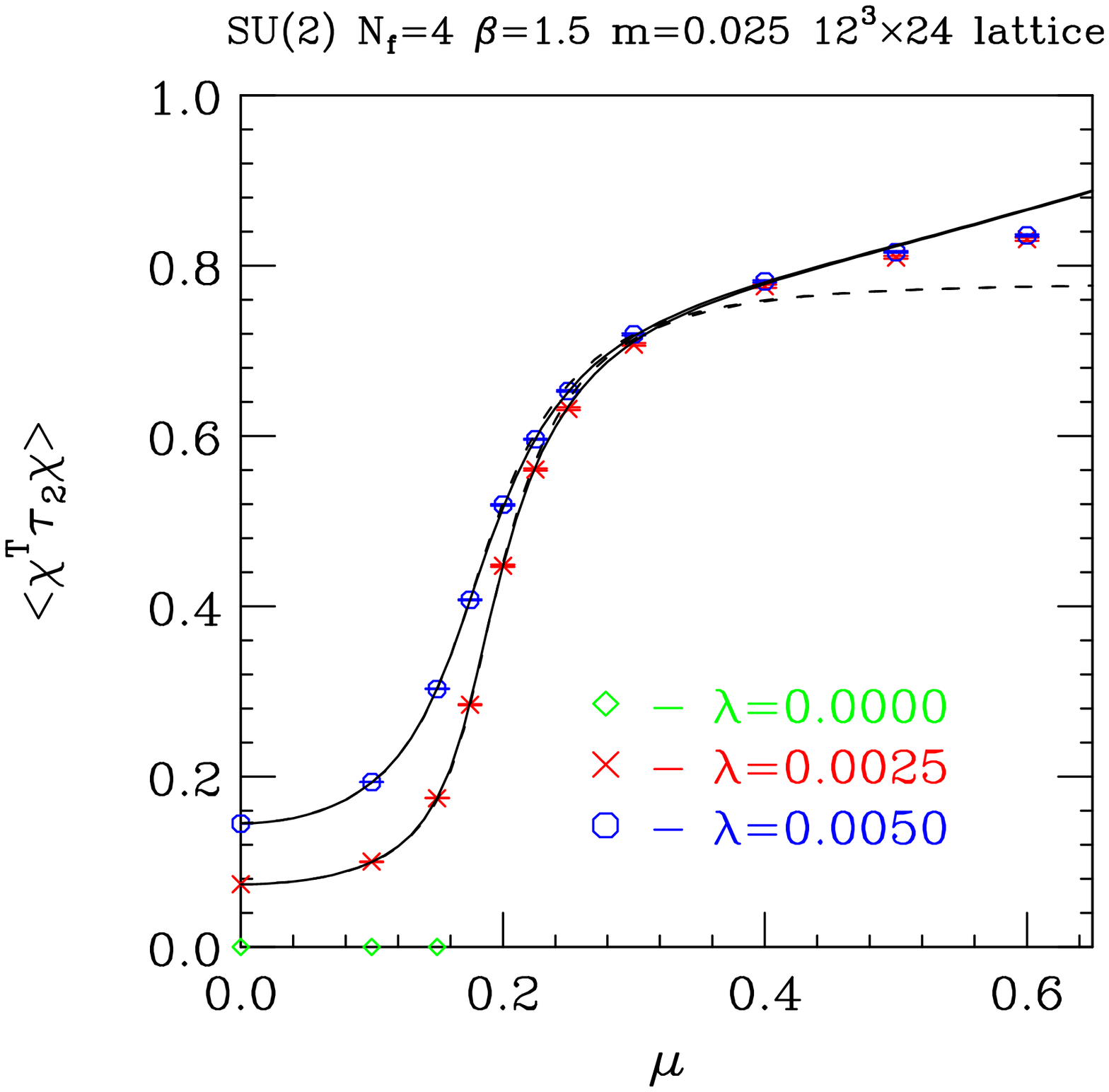}
\caption{Diquark condensate as a function of $\mu$.}
\label{fig:diquark}}
\hfill
\parbox{\halftext}{\includegraphics[width=6.6cm]{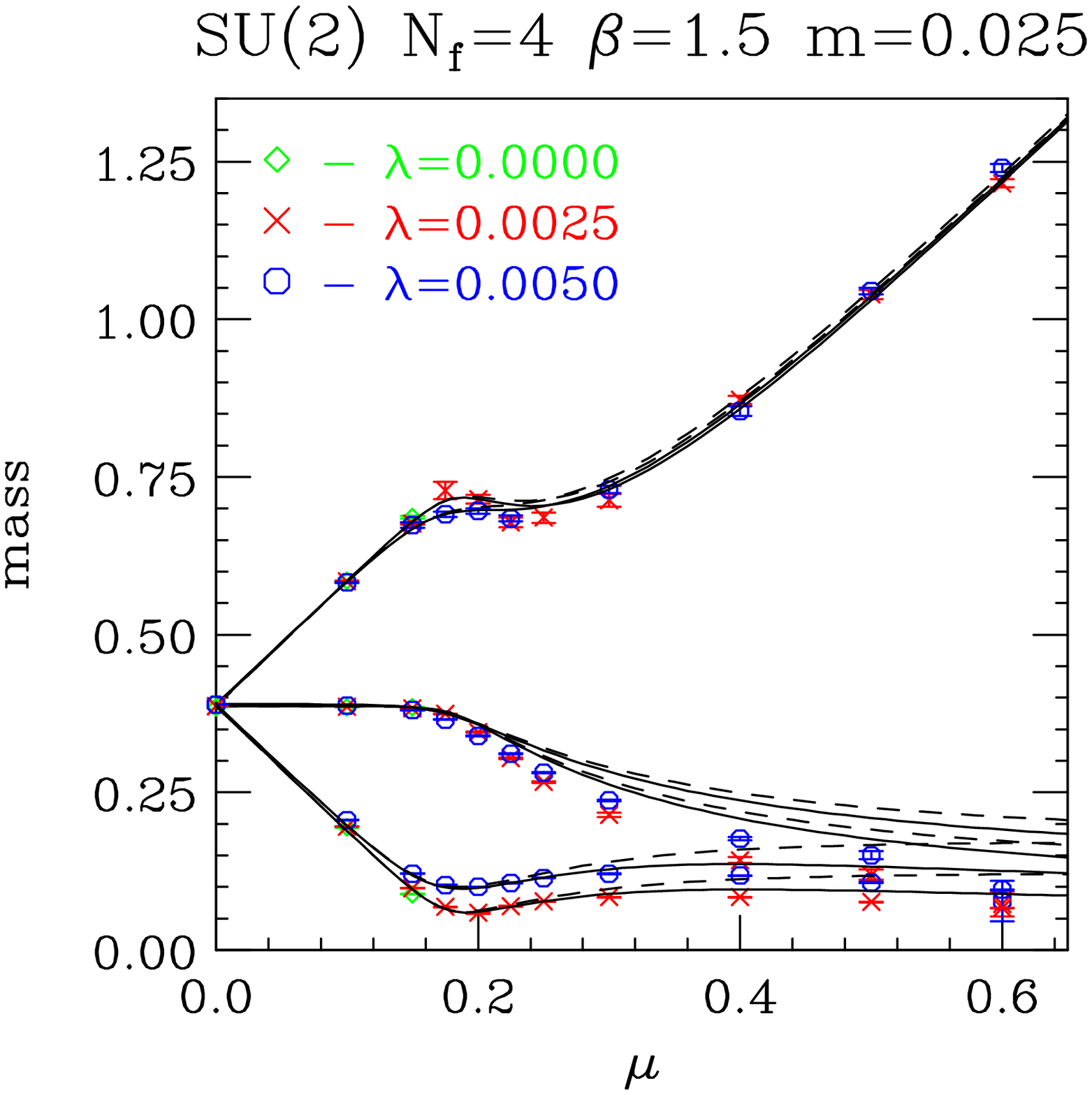}
\caption{Pseudo-Goldstone masses as functions of $\mu$.}
\label{fig:masses}}
\end{figure}

At $\lambda=0$ this theory has a phase transition at $\mu=m_\pi/2$ to a
superfluid phase with a diquark condensate which spontaneously breaks
quark-number. There is 1 true Goldstone boson. As for QCD at finite $\mu_I$,
the condensates and quark-number density are well described by a tree-level
Lagrangian of the linear sigma model class. Figure~\ref{fig:diquark} shows
the diquark condensate, which is an order parameter for this transition as a 
function of $\mu$ and $\lambda$, as measured in simulations on a 
$12^3 \times 24$ lattice at $\beta=4/g^2=1.5$ and $m=0.025$.

We measure the $\mu$ and $\lambda$ dependence of the 3 pseudo-Goldstone
bosons and compare them with the predictions of the linear sigma model
effective Lagrangian. ($m_\pi$ is small enough at $m=0.025$ for us to expect
chiral perturbation theory to be valid.) Figure~\ref{fig:masses} shows the
3 (pseudo)-Goldstone boson masses as functions of $\mu$. Although there is
good qualitative agreement, quantitative agreement is lacking once $\mu$ is
appreciably greater than $m_\pi/2$.

\section{$\chi$QCD at finite $\mu$ and T.}

We have formulated QCD with additional irrelevant chiral 4-fermion interactions
which make the Dirac operator non-singular in the chiral limit ($\chi$QCD)
\cite{Kogut:1998rg}.
The staggered fermion part of this action is:
\begin{equation}
S_f = \sum_{sites} \left\{\bar{\chi}[D\!\!\!\!/+m
      + \frac{1}{16} \sum_i (\sigma_i+i\epsilon\pi_i)]\chi\right\}
      + \sum_{\tilde{s}}\frac{1}{8}N_f\gamma(\sigma^2+\pi^2),
\end{equation}
where $\tilde{s}$ are the sites on the dual lattice, and the sum over $i$ is the
sum over the 16 sites on the dual lattice which are 1 unit away from the given
site on the original lattice. It is just the properties which allow this
action to be used in the chiral limit, that promise to make it better behaved
than the standard staggered action at finite chemical potential.

Now consider what happens when we add a chemical potential $\mu$ in the usual
way. Since the quark propagator on a single configuration no longer behaves as
$\exp[-(m_\pi/2)|x-y|]$, but is controlled by
$(\langle\sigma\rangle = \langle\bar{\chi}\chi\rangle/\gamma)+m$ rather than 
$m$, the phase of the determinant can remain well behaved for $\mu > m_\pi/2$,
and even in the chiral limit.

Thus, provided we keep $\gamma$ from becoming too large, the Swansea-Bielefeld
expansion should be well behaved, even for $m=0$. The hope is that the phase
of the determinant will remain under control up to the critical end point.
Since keeping $\gamma$ ``small'' makes the flavour symmetry breaking bad, this
should be considered as a model. It remains to see how well it represents
QCD at finite $\mu$.

In addition, we should be able to use this action to simulate with the
magnitude of the determinant (equivalent to simulating at finite $\mu_I$) at
$m=0$ and to search for the tricritical point. Here we see that the pion
consisting of a quark from the Dirac operator and an antiquark from the
conjugate Dirac operator will be relatively heavy, which prevents it from
causing problems at small $\mu$.

Preliminary studies of this theory by Barbour, Kogut and Morrison 
\cite{Barbour:1996mx} suggest
that $\chi$QCD is indeed better behaved at finite $\mu$ than standard lattice
QCD, and are consistent with the scenario presented above.

\section{Conclusions}

QCD at finite isospin chemical potential and 2-colour QCD at finite quark-number
chemical potential have some of the properties of QCD at finite 
quark/baryon-number chemical potential. Unlike QCD at finite $\mu$, they have
positive fermion determinants, so they are amenable to standard simulation
methods.

QCD at finite $\mu_I$ has a phase transition with mean-field critical exponents
to a superfluid phase in which isospin ($I_3$) and parity are broken by a
charged pion condensate at $\mu_I=m_\pi$, as predicted by chiral perturbation
theory. The dependence of the position of the finite temperature transition 
($T_c$) on $\mu$ and on $\mu_I$ is the same for $\mu$ sufficiently small; so 
simulations at finite $\mu_I$ are a convenient way of estimating the $\mu$
dependence. We will check the agreement between our measurements and the
Swansea-Bielefeld predictions. It appears that our results are consistent with
those obtained by de Forcrand and Philipsen from simulations at imaginary
$\mu$.

Our 2-flavour simulations do not show any sign of the critical endpoint
expected for QCD at finite $\mu$ and hence $\mu_I$. Perhaps the tricritical 
point on the boundary of the pion condensed phase, where the transition 
changes from second order to first order, is a remnant of this endpoint.
Unfortunately this is in the region where we cannot relate the finite $\mu_I$
and finite $\mu$ behaviour of lattice QCD.

For 2+1 flavours Fodor and Katz were able to determine the position of the
critical endpoint at finite $\mu$ \cite{Fodor:2001au,Fodor:2001pe,Fodor:2002km}.
De Forcrand and Philipsen have made
a determination of the position of this critical endpoint for 3 flavours
by continuation from imaginary $\mu$ \cite{deForcrand:2003hx}. (Similar
determinations have been made for 4 flavours by D'Elia and Lombardo
\cite{D'Elia:2002gd}.)
The Bielefeld-Swansea collaboration have
also determined the position of the critical endpoint for 3 flavours using
their series expansions around $\mu=0$ \cite{Karsch:2003va}. 
In addition to the fact that 2+1
flavours describes the real world, 2+1 or 3 flavours have the advantage that
for small enough quark masses, the critical endpoint is at the origin. This
means that one can arrange for the critical endpoint to be as close to $\mu=0$
as one desires by tuning the quark masses, and follow its trajectory as these
masses are increased. In particular, in a 3-flavour version of QCD at finite
$\mu_I$, we should be able to arrange for the critical endpoint to occur
for $\mu_I < m_\pi$, where it should be related to the critical endpoint at 
finite $\mu$. We are currently extending our QCD at finite $\mu_I$ simulations
to 3 flavours, ignoring the fact that the theory we are simulating is not
expected to have a sensible continuum limit, and hope to determine the position
of this endpoint with good precision.

There remain some questions to be answered concerning QCD at finite $\mu_I$.
These include: Does pion condensation at finite isospin density occur when the
system is also at finite baryon-number density? If so, how large can the
baryon-number density become before pion condensation no longer occurs? 
How well does the pseudo-Goldstone spectrum of lattice QCD at finite $\mu_I$
compare with the predictions of chiral perturbation theory?

The pseudo-Goldstone spectrum of 2-colour lattice QCD at finite $\mu$ shows
semi-quantitative agreement with the predictions of effective chiral 
Lagrangians.

The $\chi$QCD action shows promise for simulations at finite $\mu$ at $T$.
It should allow determination of the critical endpoint for small (or even zero)
quark masses.

\section*{Acknowledgments}

DKS is supported by the US Department of Energy, Division of High Energy
Physics, under Contract W-31-109-Eng-38. JBK and DT are supported in part by
NSF grant NSF-PHY-0102409. DT is supported in part by ``Holderbank''-Stiftung.
These simulations were performed on the IBM SP at NERSC, the IBM SP at NPACI,
the Jazz Linux cluster at LCRC, Argonne and on Linux PCs in the Argonne HEP
Division.

\end{document}